\documentclass[a4paper,11pt]{article}
\usepackage{jheppub} 
\usepackage{array} 

\arxivnumber{2607.18669} 

\title{Production of lepton-flavor-violating scalars through resonant positive-muon annihilation on atomic electrons}

\author[a]{Jinhong Shen}
\author[a]{Youpeng Wu}
\author[a]{Zijian Wang}
\author[a]{Leyun Gao}
\author[a]{Qite Li}
\author[a]{Chen Zhou}
\author[a]{Qiang Li}

\author[d,e]{Yu Xu}
\author[d,e,f]{Xueheng Zhang}
\author[d,e,f]{Liangwen Chen}
\author[d,e,f]{He Zhao}
\author[e,f]{Ruihu Zhu}
\author[d,e,f]{Zhiyu Sun}

\affiliation[a]{
School of Physics,
Peking University,
Beijing 100871,
China
}

\affiliation[d]{
Advanced Energy Science and Technology Guangdong Laboratory,
Huizhou 516000,
China
}

\affiliation[e]{
Institute of Modern Physics,
CAS,
Lanzhou 730000,
China
}

\affiliation[f]{
School of Nuclear Science and Technology,
University of Chinese Academy of Sciences,
Beijing 100049,
China
}

\emailAdd{jinhong.shen@cern.ch}
\emailAdd{youpeng.wu@cern.ch}
\emailAdd{zijian.wang@cern.ch}
\emailAdd{leyun.gao@cern.ch}
\emailAdd{qliphy0@pku.edu.cn}

\abstract{We investigate an invisible lepton-flavor-violating scalar $\phi$ with exclusive $e-\mu$ couplings and study its resonant production via $\mu^+e^-\to\phi$ in fixed-target experiments. The resonant production rate is sensitive to the momentum distribution of bound atomic electrons, including its high-momentum tail, where relativistic effects can be important, particularly for high-$Z$ targets. We therefore employ relativistic bound-state Dirac wave functions to consistently incorporate these effects into the annihilation amplitude and production cross section. We find that atomic motion broadens the resonance lineshape in lead targets. Using the resulting cross section, we propose a HIAF experiment that can probe couplings at the $10^{-5}$ level at 90\% confidence level near resonance with $6\times10^{10}$ muons on target, corresponding to less than one day of data taking. This projected sensitivity highlights the potential of high-intensity muon fixed-target experiments as a complementary probe of lepton-flavor violation.}

\begin{document}
\maketitle
\flushbottom
\section{Introduction}
Lepton-flavor-violating (LFV) interactions provide a sensitive probe of physics beyond the Standard Model (BSM), where charged-lepton flavor is conserved at the renormalizable level, apart from tiny effects induced by neutrino masses. A simple BSM scenario introduces a hypothetical scalar mediator, $\phi$, that couples to opposite-sign electron--muon pairs~\cite{Gninenko:2022ttd,Xiang:2026}. While such interactions are tightly constrained by the nonobservation of LFV processes, they remain experimentally well motivated, as they can produce distinctive signatures ranging from rare low-energy decays to resonant production at high-intensity facilities. In contrast to observables such as $\mu \to e \gamma$~\cite{baldini2016search,yonemoto2025new}, $\mu \to 3 e$~\cite{BERTL19851}, $\mu$--$e$ conversion in muonic atoms~\cite{bertl2006search} and muonium to antimuonium conversion \cite{PhysRevLett.82.49,Bai:2026mace}, which probe the same LFV coupling through loop-induced corrections or off-shell mediator exchange, the $\mu^+e^-\to\phi$ process directly accesses the LFV vertex via resonant on-shell scalar production. It therefore provides a complementary and more direct probe of LFV dynamics. 

The $\mu^+ e^- \to \phi$ events can be efficiently produced using high-intensity positive-muon beams incident on material targets. Owing to the strong resonance enhancement, the experiment is highly sensitive to $m_\phi$ values near the center-of-mass energy of the $\mu^+ e^-$ system, which can significantly exceed those accessible in current experiments. The PKMu experiment~\cite{Gao:2025lbx,Liu:2025vyq} performed a model-independent search for elastic muon-philic dark matter using a simple yet effective experimental facility~\cite{Yu:2024spj,liu2026probing}. Assuming that the produced $\phi$ is invisible to the detector, the PKMu tracking system can be configured to detect the incoming muon while vetoing the subsequent presence of energetic muons or electrons. The high-intensity GeV-scale positive-muon beams expected to be available at the near future HIAF facility~\cite{An:2025lws} will provide the sensitivity to scalar masses around $m_\phi \sim 10^{2}~\mathrm{MeV}$.

In fixed-target experiments, the target electrons are bound and possess nontrivial momentum distributions. For high-Z targets, their momenta can reach relativistic scales of $\mathcal{O}(10~\mathrm{MeV})$, rendering the commonly adopted electron-at-rest approximation inadequate. Convolution of the resonant cross section with the electron momentum distribution broadens the resonance into a material-dependent lineshape and can even allow production beyond the nominal free-electron threshold through the high-momentum tail of the bound-state wave function \cite{Arias-Aragon:2024qji,Arias-Aragon:2025xcc,Luc:2025,DLSSW}.

Existing studies incorporate atomic effects through Compton-profile convolutions \cite{Arias-Aragon:2024qji} or perturbative corrections to free-scattering amplitudes~\cite{Plestid:2024}. In this work, we perform a first-principle calculation based on relativistic atomic Dirac wave functions, enabling a consistent treatment of bound-electron effects at both the amplitude and cross-section levels. We find explicit deviations from the electron-at-rest approximation, while maintaining consistency with the Compton-profile approach in the kinematic regime relevant to current high-intensity muon-beam facilities. We then evaluate the sensitivities of proposed LFV scalar mediator experiments on HIAF and derive projected constraints on its coupling to $\mu^+e^-$. The same methodology can be applied to other processes involving bound electrons, such as precision muon-electron scattering at MUonE \cite{Ignatov:2023wma,abbiendi2022status}.

\section{Formalism}
At the level of an effective theory, we take the interaction in the charged-lepton mass basis to be
\begin{equation}
  \mathcal{L}_{\rm int}
  = - g_\phi\, \phi\, e\,\bar\mu
  + \text{h.c.},\qquad
  \mathcal{L}_{\rm dark} = -g_D\,\phi\,\bar{\chi}\,\chi
\end{equation}
where $g_\phi$ is the off-diagonal LFV Yukawa coupling, and $g_D$ is the dark-sector coupling.
For $m_\phi > m_\mu + m_e \approx 106.2$~MeV, the decay $\phi \to e\mu$ is
kinematically open.

\begin{figure}
    \centering
\includegraphics[width=0.6\linewidth]{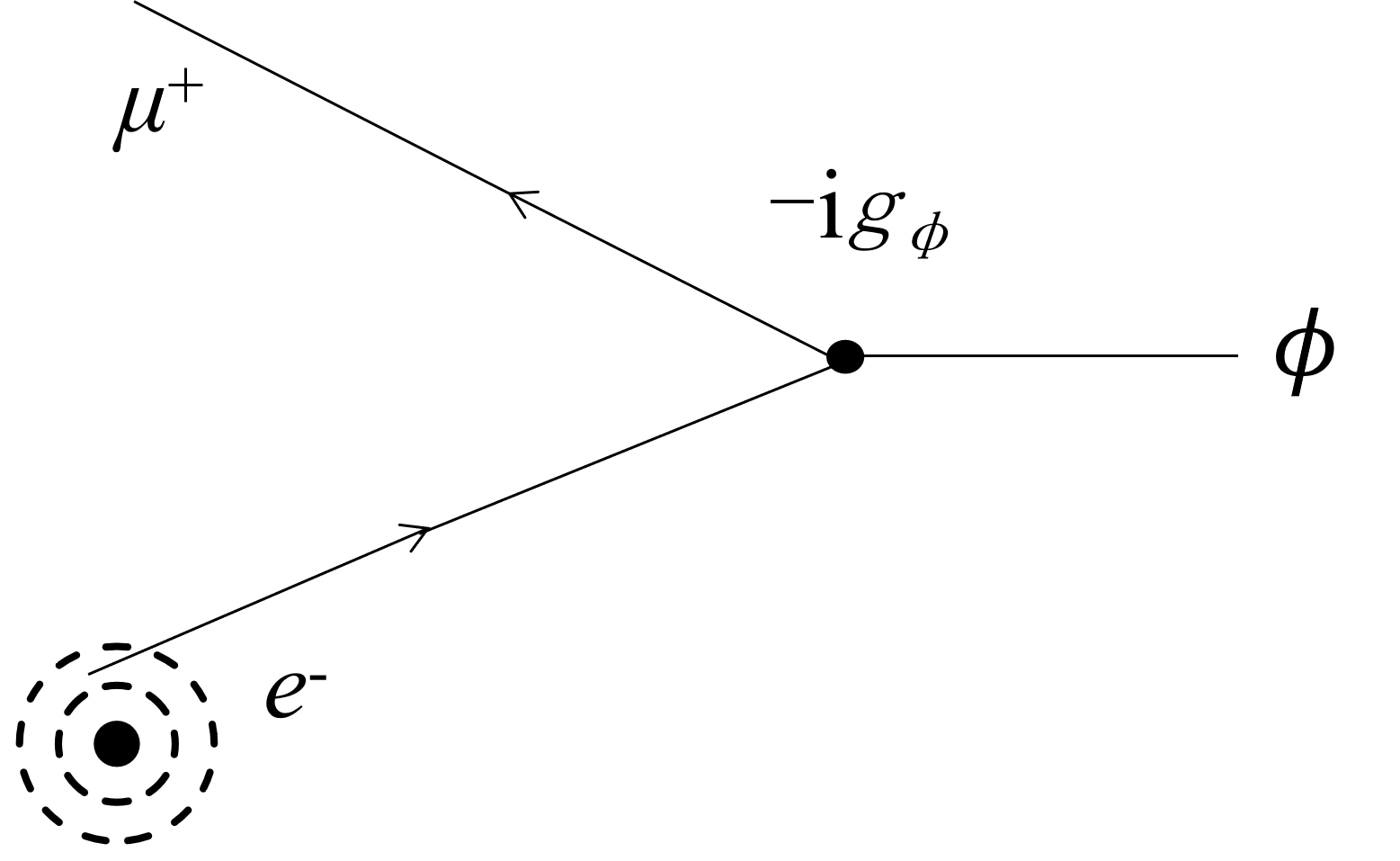}
    \caption{The Feynman diagram for the lepton-flavor-violating (LFV) interaction studied in this work: an incoming $\mu^+$ annihilates with a bound electron to produce an invisible scalar media $\phi$.  The coupling strength is denoted by $g_\phi$. }
    \label{fig:Feynman}
\end{figure}

This paper adopts the convention that all other scalar couplings to charged leptons are negligible, so that $\phi$ interacts only through the $e$--$\mu$ flavor-changing channel. The Feynman diagram of the process $\mu^+ e^- \to \phi$ is depicted in figure \ref{fig:Feynman}. 

In this work we focus on the invisible decay signature of the resonantly produced scalar,
\begin{equation}
  \mu^+ e^- \to \phi \to \chi\bar{\chi}.
\end{equation}
Although the visible LFV decay channel $\phi \to e\mu$ is open for $m_\phi > m_\mu + m_e$, it is subdominant in the parameter region considered here. For $g_D > g_\phi$, the invisible width dominates over the LFV width, as the corresponding partial widths scale as $g_D^2$ and $g_\phi^2$. For the benchmark couplings $g_\phi=10^{-5}$ and $g_D=10^{-4}$, the typical visible and invisible widths are $\Gamma_{\rm vis}\sim 10^{-11}\,\mathrm{MeV}$ and $\Gamma_{\rm inv}\sim 10^{-8}\,\mathrm{MeV}$. The resulting correction from a finite invisible branching ratio is negligible, and we therefore set
$\mathrm{Br}(\phi\to\chi\bar{\chi}) \simeq 1$ throughout the analysis. 

Our calculations are carried out based on the original formalism presented in Ref \cite{Arias-Aragon:2024qji}, which was derived for Compton profiles and implemented with free-particle scattering amplitudes. We further use relativistic atomic Dirac spinors leveraging the techniques from~\cite{Gardner:1994} as will be discussed in~\cite{DLSSW}.

\subsection{Dirac Spinors}
The Dirac spinors of the relativistic bound state electron and free muon in momentum representation are as follows.
\[	e^-: \mathcal{U}_{E\kappa m}(K_A,m),\quad \mu^+:v_B\left(P_B,s \right)
\]
Here $E$ denotes the electron energy. $m$ denotes the magnetic quantum number. \(\kappa=(l-j)(2j+1)\) denotes the Dirac angular quantum number, where $l$ and $j$ are the orbital and total angular momentum quantum numbers. $K_A$ and $P_B$ denote the four-momenta of the electron and muon respectively and $s$ labels the spin state of the incoming $\mu^+$.

$v_B $ takes the form \cite{Peskin:1995page48}:
\begin{equation}
    v_B=\left[
        \sqrt{\mathbf{p_B\cdot\boldsymbol{\sigma}}}\,\xi^s,
\sqrt{\mathbf{p_B}\cdot\bar{\boldsymbol{\sigma}}}\,\xi^s
   \right]^T ,\quad s=1,2
\end{equation}
$\xi^1=(1,0)$ , $\xi^2=(0,1)$. $\boldsymbol{\sigma}$ denotes the vector of Pauli matrices. $\mathbf{p_B}$ denotes the muon three-momentum.

The bound electron Dirac spinor $\mathcal{U}_{E\kappa m} $ takes the form \cite{Gardner:1994}:
\begin{equation}
\label{eq:f_and_g}
    \mathcal{U}_{E\kappa m}=\frac{4\pi}{k_A}(-i)^l\begin{bmatrix}
    g_{E\kappa}\mathcal{Y}_{+\kappa m}(\mathbf{\hat{k}_A})\\
    -f_{E\kappa}\mathcal{Y}_{-\kappa m}(\mathbf{\hat{k}_A}) 
\end{bmatrix} 
\end{equation}
The angular dependence is described by the spinor spherical harmonics $\mathcal{Y}{\pm\kappa m}$, while the radial dependence is encoded in the large and small components, denoted by $g_{E\kappa}$ and $f_{E\kappa}$, respectively. Here $\mathbf{\hat{k}_A}$ denotes the direction of the electron momentum $\mathbf{k_A}$. We adopt the following normalization convention:
\begin{equation}
   \frac{1}{(2\pi)^3} \sum_m\int \mathcal{U}^\dagger _{E\kappa m}\mathcal{U}_{E\kappa m} \,d^3k_A=2E_A
\end{equation}
Here $E_A$ denotes the electron energy.

\subsection{Cross Section}
The characteristic timescale of the $\mu e$ annihilation is set by the interaction of the incoming muon with the entire atom, leading to the normalization flux factor $1/|v_B|$. 

Enforcing energy and momentum conservation, the total cross section summing over all atomic electrons is given as follows.
\begin{equation}
    \label{eq:cross_section}
    \sigma = \sum_{Z}\sigma_{E\kappa}=\sum_{Z}\int \frac{d^3k_A}{(2\pi)^3}\frac{\left|\mathcal{M}\right|_\kappa^2}{2E_B\left|v_B\right|}d\Phi,
\end{equation}
\begin{equation}
\label{eq:integration variable}
    d\Phi = (2\pi)^4 \delta \left(E_A + E_B -  E_\phi\right)\delta^{(3)}\left(\mathbf{k_A}+\mathbf{p_B}-\mathbf{p_\phi}\right)  \frac{d^3p_\phi}{(2\pi)^32E_\phi}.
\end{equation}
$\sigma_{E\kappa}$ denotes the cross section for an electron with energy $E$ and Dirac angular quantum number $\kappa$. $|\mathcal{M}|_{\kappa }^2$ denotes the corresponding squared Feynman invariant amplitude. $Z$ denotes the atomic number of the target atom.

\subsection{Feynman Invariant Amplitude}
The Feynman invariant amplitude $\mathcal{M}_{\kappa m}$ takes the form:
\begin{equation}
    \mathcal{M}_{\kappa m}=g_\phi\times \bar{v}\left( P_{B},s  \right) \times\mathcal{U}_{E\kappa m}(K_A,m).
\end{equation}
The squared Feynman invariant amplitude is obtained by averaging over the initial-state spins and summing over all final-state spins: 
    \begin{equation}
    \label{eq:Amplitude}
	\overline{|\mathcal{M}|_\kappa^2}
	=\frac{1}{2}g_\phi^2 \sum_{s} [\bar{v}(P_B,s)v(P_B,s)]\sum_{m}{\left[\mathcal{U}_{E\kappa m}\left( K_A,m \right) \bar{\mathcal{U}}_{E\kappa m}\left( K_{A},m \right) \right]}.
\end{equation}
The summations are performed respectively:
\begin{equation}
\label{eq:Dirac spinor of muon} 
    \sum_{s} [\bar{v}(P_B,s)v(P_B,s)]= /\!\!\!P_{B}-m_{B} ,
\end{equation}
\begin{equation}
\label{eq:Dirac function for bound electron}
\sum_{m}{\left[\mathcal{U}_{E\kappa m}\left( P_A,m \right) \bar{\mathcal{U}}_{E\kappa m}\left( P_{A},m \right) \right]}
  =\frac{2(2j+1)\pi}{k_A^2}
   \begin{bmatrix}
       g^2&\,-gf(\boldsymbol{\sigma}\cdot\mathbf{\hat{k}_A})\\
       gf(\boldsymbol{\sigma}\cdot\mathbf{\hat{k}_A})&\, -f^2
   \end{bmatrix},\\
\end{equation}
The subscripts of f and g function from Eq \eqref{eq:f_and_g} are omitted in simplicity.

Combining Eqs.~\eqref{eq:Amplitude}, \eqref {eq:Dirac spinor of muon} and \eqref{eq:Dirac function for bound electron}, the squared amplitude takes the form:
\begin{equation}
  \overline{|\mathcal{M}|_\kappa^2}
	=\frac{(2j+1)\pi g_\phi^2}{k_A^2}Tr\left\{\left( /\!\!\!P_{B}-m_{B} \right)
    \begin{bmatrix}
       g^2&\,-gf(\boldsymbol{\sigma}\cdot\mathbf{\hat{k}_A})\\
       gf(\boldsymbol{\sigma}\cdot\mathbf{\hat{k}_A})&\,-f^2
   \end{bmatrix}\right\}.\\  
\end{equation}
Calculating the trace:
\begin{equation}
\label{eq:final amplitude}
   \overline{|\mathcal{M}|_\kappa^2}
   =\frac{2(2j+1)\pi g_\phi^2}{k_A^2} \left[
    g^2(E_B-m_B)-2gf(\mathbf {p_B}\cdot\mathbf{\hat{k}_A})+f^2(E_B+m_B)
   \right].
\end{equation}

\subsection{Phase Space Integration}

Combining Eqs. \eqref{eq:cross_section} \eqref{eq:integration variable} and \eqref{eq:final amplitude}, the cross section is given by:
\begin{equation}
    \sigma=\sum_{Z}
    \int\frac{\ (2j+1)g_\phi^2\left[
    (E_B-m_B)g^2-2gfp_Bx_0+f^2(E_B+m_B)]
   \right] }{8 p_B^2 E_A k_A}{d}k_A
\end{equation}
where 
\begin{equation}
    x_0\left( k_A \right) =\frac{E_{\phi}^{2}-m_{\phi}^{2}-k_{A}^{2}-p_{B}^{2}}{2 k_A p_B}.
\end{equation}
Here $x_0$ denotes the cosine value of the angle subtended between the muon momentum and the electron momentum. The integration phase space covers the region where $x_0(k_A)\in[-1,1]$.





A lead target is considered in this work. The electronic configuration of a lead atom is $[\mathrm{Xe}](4f)^{14}(5d)^{10}(6s)^2(6p)^2 .$ Our calculation includes all occupied relativistic subshells of the lead atom. The incoming beam particle is treated as an external scattering state, and the resonant annihilation rate is evaluated from the overlap with the bound-electron Dirac wave functions in momentum space. Consequently, all occupied electrons contribute to the inclusive atomic cross section. The total cross section is therefore obtained as
\begin{equation}    
\sigma_{\mathrm{Pb}} = \sum_{Z}\sigma_{n\kappa}=\sum_{n\kappa}^{\mathrm{occ.}}
N_{n\kappa}\, \sigma_{n\kappa},
\end{equation}
where \(N_{n\kappa}\) denotes the occupation number of the subshell \(n\kappa\). For neutral lead, \(\sum_{n\kappa}^{\mathrm{occ.}} N_{n\kappa}=82.\) This full-shell summation is consistent with the momentum-distribution folding approach, in which the electron momentum density is normalized to the total electron number \(Z=82\).

We compare our results with those obtained using two representative treatments of atomic effects. The at-rest (Free/rest) approximation neglects the electron momentum and evaluates the cross section using free-particle Dirac spinors. In this method the squared Feynman invariant amplitude $|\mathcal{M}|^2= 2g_\phi^2(E_AE_B-m_Am_B)$. The Compton profile approach\cite{Arias-Aragon:2024qji} partially incorporates bound-state effects through the number density of atomic electrons $n(k_A)$ encoded in the target's Compton profile. But the squared Feynman invariant amplitude ($|\mathcal{M}|^2= 2g_\phi^2(E_AE_B-k_Ap_Bx_0-m_Am_B)$) it employs is still derived from free-particle spinors. 

\begin{figure}[htbp]
    \centering
    \includegraphics[width=0.8\linewidth]{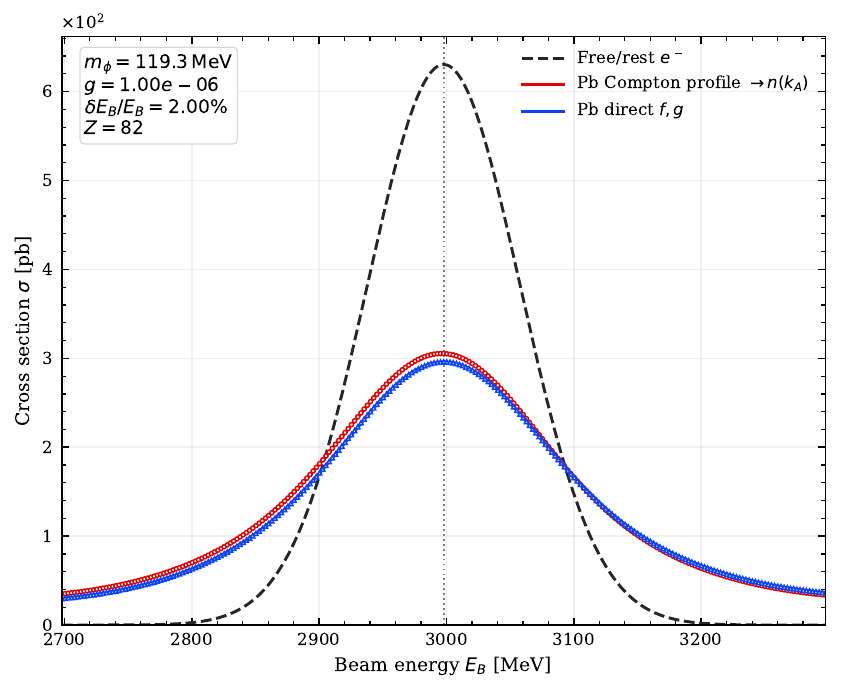}
    \caption{Comparison of the resonant production cross section obtained using three different treatments of atomic effects for a lead target ($Z=82$), assuming $m_\phi=119.3~\mathrm{MeV}$, $g_\phi=10^{-6}$, center energy $E_B=3GeV$ and a beam-energy spread of $\delta_{E}/E=2\%$ \cite{Xu:2025hiaf}. Atomic effects significantly broaden the resonance and reduce the peak cross section relative to the free-electron approximation. The f,g method predicts a lower peak than the Compton Profile method at the center energy, as the modification of the Feynman invariant amplitude by the bound-electron momentum distribution further smears the resonance. The slight asymmetry originates from the asymmetry of f,g functions.}
    \label{fig:upper limit}
\end{figure}

Figure \ref{fig:upper limit} shows the resonant production cross section obtained using the aforementioned three different treatments of atomic effects for a lead target.  The mass of the scalar $\phi$ is fixed at the resonant mass $m_\phi=\sqrt{m_A^2+m_B^2+2E_B m_A}$ , where $k_A=0$. The black dashed curve corresponds to the conventional free-electron (at-rest) approximation. The red solid curve is obtained by convolving the free-electron cross section with the bound-electron number density $n(k_A)$ derived from the Compton profile of Pb \cite{Biggs:1975}, following the method of \cite{Arias-Aragon:2024qji}. This method retains the free-particle Feynman invariant amplitude. The blue solid curve shows the result of this work in which the bound electron is treated consistently using $f$ and $g$ functions in both the squared Feynman amplitude and the phase space integration. The $f$ and $g$ functions are computed from the Roothan-Hartree-Fock (RHF) wave functions \cite{huzinaga1965gaussian,STEINBORN19731,clementi1974roothaan,KAIJSER197737,PhysRevA.18.1,Maretis1979,Weniger1983,Weniger1985}.

\section{Experimental Setup and Simulation Results}

\subsection{Experimental Setup}

The proposed experiment is designed to run at the High Intensity 
heavy-ion Accelerator Facility (HIAF)~\cite{Zhou:2022hiaf}, currently under construction at the Institute of Modern Physics (IMP) in Huizhou, China. 
The muon beam is produced and transported by the High energy FRagment Separator (HFRS) beamline~\cite{Xu:2025hiaf}, which is 192~m long with a maximum magnetic rigidity of 25~Tm, enabling the transport of muons with momenta up to 7.5~GeV/$c$. Using a 9.3~GeV proton primary beam incident on a graphite target, pions are produced and subsequently decay in flight along the beamline; the total HFRS length is well-matched to the decay length of a $\sim$3~GeV/$c$ pion ($\approx 170$~m). A two-section magnetic rigidity purification strategy raises the $\mu^+$ beam purity to 100\% at an intensity of $\mathcal{O}(10^5)~\mu^+/\mathrm{s}$~\cite{Xu:2025hiaf}. For the present study, we adopt the benchmark parameters $E_B = 3$~GeV, an intensity of $7.5\times 10^5~\mu^+/\mathrm{s}$, and a transverse beam spot of $\sigma_x = \sigma_y = 30$~mm.

Crucially, the resonance condition $\sqrt{s}=m_\phi$ for $\mu^+e^-\to\phi$ on atomic electrons at rest maps a 3~GeV muon beam onto scalar masses in the range 100--200~MeV---precisely the window where no competitive experimental limit currently exists, making HIAF uniquely suited for this search.

\begin{figure}[htbp]
    \centering
    \includegraphics[width=0.8\linewidth]{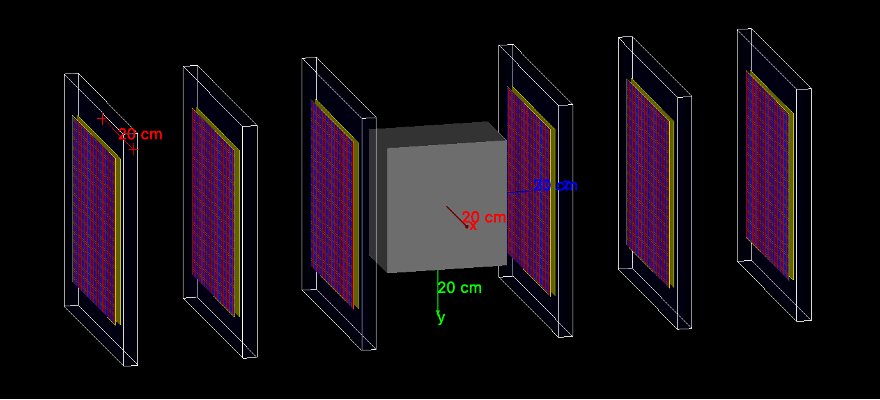}
    \caption{Schematic view of the RPC detector system at HIAF. Six RPC 
planes are arranged symmetrically along the beam axis ($z$-direction), 
with three upstream and three downstream of the lead target (grey block). 
The red and blue grid lines on each RPC face represent the $x$- and 
$y$-readout strips, respectively. The minor inter-RPC gap is 170~mm; 
the major gap surrounding the target is 300~mm. Axes follow the 
convention: $z$~along the beam direction, $y$~vertical, and $x$~horizontal. 
The geometry is implemented in \textsc{Geant4} using the \texttt{FTFP\_BERT} 
physics list.}
    \label{fig:detector}
\end{figure}

The detector system, illustrated in Fig.~\ref{fig:detector}, consists of six Resistive Plate Chamber (RPC) modules arranged symmetrically along the beam axis: three upstream and three downstream of a lead target. The target dimensions are $120\text{~mm} \times 210\text{~mm} \times 200\text{~mm}$, providing sufficient material for resonant production while remaining compact enough to preserve the minimalist signal topology. Adjacent RPC planes are separated by a minor gap of 170~mm, while the upstream and downstream groups are separated by a major gap of 300~mm centered on the target. Each RPC has a sensitive area of $280\text{~mm} \times 280\text{~mm}$ and employs a 2S module readout scheme with delay-line position reconstruction via energy-weighted averaging per layer.

\subsection{Monte Carlo Simulation}

Signal and background processes are simulated within the \textsc{Geant4} 
framework using the \texttt{FTFP\_BERT} physics list, supplemented by 
89~additional background processes covering the dominant muon interaction 
channels: $\mu$ decay, $\mu e$ Coulomb scattering, $\mu$ bremsstrahlung, 
$\mu$ ionisation, $\mu^-$ nuclear capture, $\mu e$ pair production, 
$\mu$-nuclear interaction, multiple Coulomb scattering, as well as neutron 
radiative capture, neutron inelastic scattering, and hadronic elastic 
scattering. The signal process $\mu^+ e^- \to \phi$ is implemented with 
$\phi$ decaying invisibly; the electron number density and mean free 
path are computed at each simulation step to correctly account for the 
production rate in the target material. Signal events are generated at 
a reference coupling $g_\phi^{\mathrm{ref}} = 10^{-6}$ and subsequently 
reweighted to arbitrary coupling values via cross-section scaling. A total 
of $2.5 \times 10^7$ events are generated. RPC hits are reconstructed 
using a delay-line readout simulation in which the hit position per layer 
is determined by energy-weighted averaging.

The signal topology exploits the fully invisible final state of $\phi$: 
a muon enters the upstream RPC stack, annihilates in the target, and 
produces no downstream activity. We define the following four sequential 
selection criteria:
\begin{itemize}
    \item[]\textbf{Cut 1.} \textbf{Upstream coincidence}: all three 
    upstream RPC planes register a hit, confirming a well-reconstructed 
    incoming muon track.
    \item[]\textbf{Cut 2.} \textbf{Downstream veto}: no hit is recorded 
    in any of the three downstream RPC planes, consistent with an 
    invisible final state.
    \item[]\textbf{Cut 3.} \textbf{Beam-spot containment}: the hit 
    position in the first upstream RPC layer satisfies $r < 30$~mm, 
    corresponding to approximately $1\sigma$ of the beam spot, ensuring 
    the muon originates from the beam core.
    \item[]\textbf{Cut 4.} \textbf{Target pointing}: the incident 
    track, extrapolated from the upstream hits, is required to intersect 
    the fiducial volume of the lead target.
\end{itemize}

The resulting cut-flow is summarized in Table~\ref{tab:cutflow}. 
After all four cuts, $45{,}969$ signal events survive from an 
initial $221{,}983$, while the background is reduced from 
$24{,}769{,}782$ to $387$ events---a rejection factor exceeding 
$99.999\%$. The signal efficiency remains at approximately $20.7\%$; 
the dominant loss arises from Cut~3, which preferentially removes 
signal muons that scatter laterally in the upstream material before 
reaching the first RPC layer. Cut~2 provides the most powerful 
background rejection, suppressing the background by more than four orders of magnitude by vetoing all events with downstream activity.

\begin{table}[htbp]
\centering
\caption{Cutflow summary for signal and background events.}
\label{tab:cutflow}
\begin{tabular}{lrrrrr}
\hline
\textbf{Process} & \textbf{Generated} & \textbf{Cut 1} 
    & \textbf{Cut 2} & \textbf{Cut 3} & \textbf{Cut 4} \\
\hline
Signal($\times10^{-4}$)& $221{,}983$ & $221{,}983$ & $219{,}061$ 
           & $46{,}108$  & $45{,}969$ \\
Background & $24{,}769{,}782$ & $24{,}764{,}801$ & $2{,}211$ 
           & $395$ & $387$ \\
\hline
\end{tabular}
\end{table}

Assuming no signal observation, we derive projected 90\% confidence 
level upper limits on the LFV coupling $g_\phi$ as a function of 
the scalar mass $m_\phi$. The sensitivity is evaluated for several 
benchmark running durations at the nominal beam intensity of 
$7.5 \times 10^5~\mu^+/\mathrm{s}$.

\begin{figure}[htbp]
    \centering
    \includegraphics[width=0.85\linewidth]{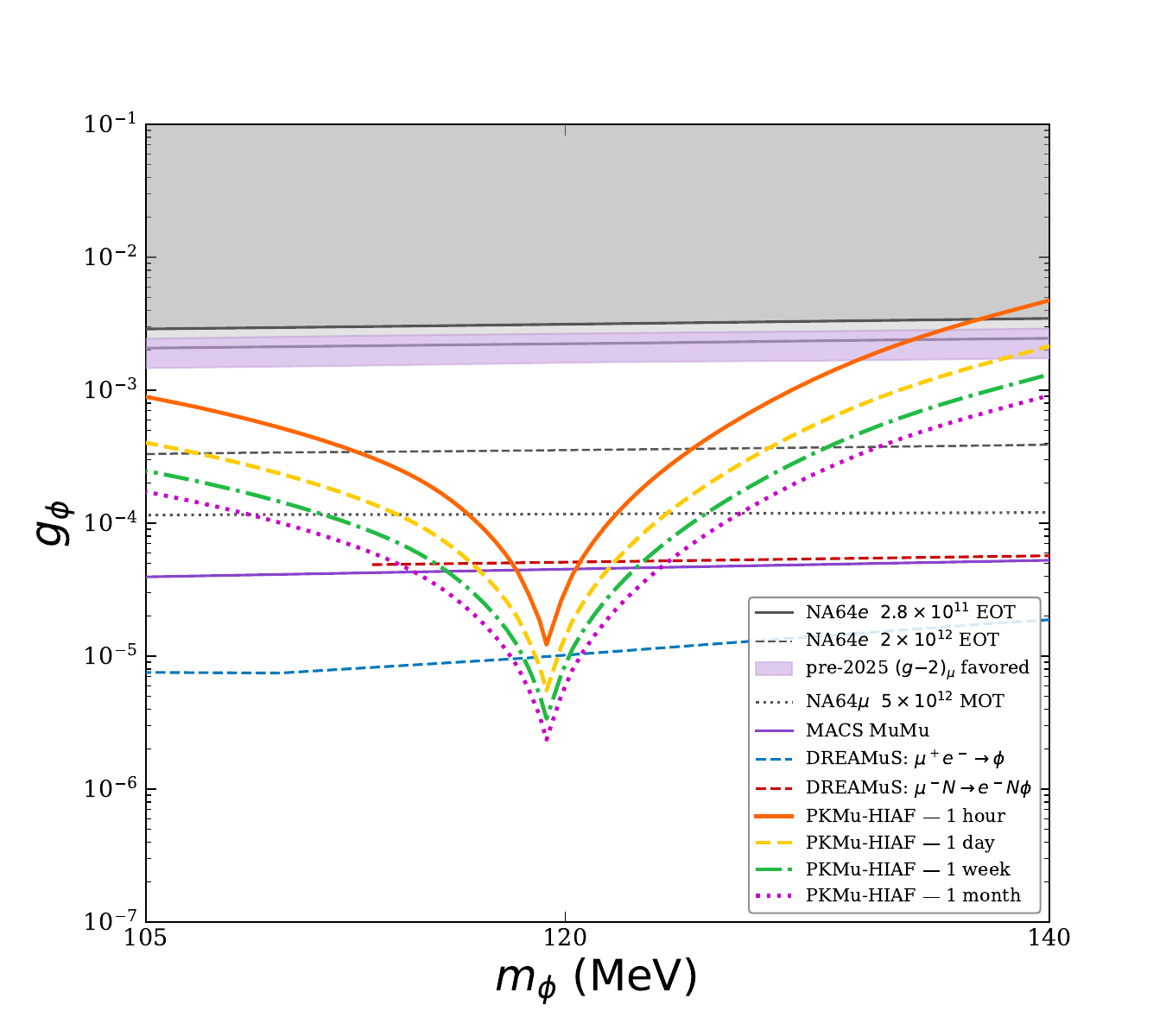}
    \caption{Projected 90\% confidence level upper limits on the 
lepton-flavour-violating coupling $g_\phi$ as a function of the 
scalar mass $m_\phi$, for four benchmark running durations at the 
HIAF-PKMu experiment (coloured curves) with fixed beam energy 3 GeV. The grey shaded region is 
excluded by NA64$e$~\cite{Gninenko:2022ttd}, with the solid and dashed grey 
lines indicating the limits at $2.8\times10^{11}$ and 
$2\times10^{12}$ EOT, respectively. The purple band shows the 
parameter space favoured by the pre-2025 $(g{-}2)_\mu$ 
anomaly~\cite{g2mu}. The dotted grey line shows the projected 
sensitivity of NA64$\mu$ at $5\times10^{12}$ 
MOT~\cite{Cesarotti:2023nep}. The violet solid line indicate the MACS $M(\mu^+e^-)\bar M(\mu^-e^+)$ conversion result\cite{PhysRevLett.82.49}. The blue and red dashed lines 
indicate the DREAMuS projections for the $\mu^+e^-\to\phi$ and 
$\mu^-N\to e^-N\phi$ channels, respectively~\cite{Xiang:2026}. 
All limits assume no signal observation.}
  \label{fig:Upperlimit}
\end{figure}

As shown in Fig.~\ref{fig:Upperlimit}, the proposed experiment 
achieves dominant sensitivity in the 100--200~MeV 
mass window---a region currently unconstrained by any existing 
search~\cite{Gninenko:2022ttd,PhysRevLett.82.49,Cesarotti:2023nep}. With only 10 minutes of data-taking 
($4.5 \times 10^8~\mathrm{MOT}$), the projected limit already 
surpasses the full-dataset result of NA64$\mu$~\cite{Cesarotti:2023nep}. 
After 1 hour ($2.7 \times 10^9~\mathrm{MOT}$), the sensitivity 
rivals the 2000-hour projection of DREAMuS~\cite{Xiang:2026}. 
Extended running of one day or one year yields correspondingly 
stronger constraints, with the coupling reach approaching 
$g_\phi \sim 10^{-5} - 10^{-6}$ at 90\% CL across the resonance window. 
These projections establish the HIAF-PKMu experiment as the 
world-leading probe of $e$--$\mu$ LFV scalar interactions in 
this mass range.

\section{Systematic Uncertainty Analysis}

The sensitivity projections in Sec.~3 are obtained assuming a nominal detector response. 
In particular, the residual background yield after the downstream-veto selection is evaluated using the central RPC efficiency adopted in the simulation, $\epsilon_{\rm RPC,0}=95\%$ per plane. 
We therefore separate the impact of the nominal RPC inefficiency from the additional degradation associated with the uncertainty in its calibration. 
A data-driven calibration strategy for the RPC efficiency is described in Appendix~\ref{sec:calibration}. 

In the projection, the calibration uncertainty is modelled by a single nuisance parameter $\theta_R$ (constrained by a unit Gaussian prior) that shifts the per-plane RPC efficiency as $\epsilon_{\rm RPC}(\theta_R)=\epsilon_{\rm RPC,0}+\delta\epsilon_{\rm RPC}\theta_R$. 
The three downstream RPC planes share a common nuisance parameter, corresponding to a fully correlated calibration uncertainty. This choice is conservative, since a common downward fluctuation maximises the triple-miss probability. 
The residual background $B_R(\theta_R)$ denotes the yield after the downstream RPC veto, defined in Appendix~\ref{sec:calibration} as $B_0\bigl[(1-\epsilon_{\rm RPC}(\theta_R))/(1-\epsilon_{\rm RPC,0})\bigr]^3$, where $B_0$ is the nominal residual background yield at $\epsilon_{\rm RPC,0}$, scaled to the exposure considered.

\subsection{Scintillator Detector Veto}

A further reduction of the detector-induced invisible background can be achieved by placing a plastic scintillator layer behind the downstream RPC stack. We assume a scintillator detection efficiency $\epsilon_S = 99\%$ for minimum-ionising particles. Let $h_i=1$ denote a hit in the $i$-th downstream RPC plane and $h_i=0$ otherwise. The three RPC planes are combined into a binary variable $R$, and the scintillator response is denoted by $S$:
\begin{equation}
  R = \begin{cases}
    0, & h_1 = h_2 = h_3 = 0, \\
    1, & \mathrm{otherwise},
  \end{cases}
  \qquad
  S = \begin{cases}
    0, & \mathrm{scintillator\ has\ no\ hit}, \\
    1, & \mathrm{scintillator\ hit}.
  \end{cases}
\end{equation}
The signal region and three auxiliary control regions are then defined as
\begin{equation}
  \begin{aligned}
    \mathrm{SR}  &: R = 0,\; S = 0, \qquad
    \mathrm{CR01} &: R = 0,\; S = 1, \\
    \mathrm{CR10} &: R = 1,\; S = 0, \qquad
    \mathrm{BKG} &: R = 1,\; S = 1.
  \end{aligned}
\end{equation}
In an analysis with real data, these four regions allow the residual background to be constrained via an ABCD method,
\begin{equation}
  B_{00} = \kappa \, \frac{B_{01}\,B_{10}}{B_{11}},
\end{equation}
where $B_{ij}$ denotes the background yield in the region with $R=i$ and $S=j$, and $\kappa$ parametrizes residual correlations between the RPC and scintillator responses. The ABCD relation is not used in the numerical projections shown below;
it only illustrates a possible data-driven validation strategy for a
future analysis. No closure uncertainty associated with $\kappa$ is
included in Fig.~\ref{fig:Uncertainty}.

The scintillator efficiency is parameterised with a nuisance parameter analogously to the RPC efficiency,
\begin{equation}
  \epsilon_S(\theta_S) = \epsilon_{S,0} + \delta\epsilon_S\,\theta_S,
\end{equation}
with $\epsilon_{S,0}=99\%$ and $\delta\epsilon_S=0.1\%$. The combined background and signal yields in the scintillator-assisted configuration are
\begin{equation}
  B_{R+S}(\theta_R,\theta_S) = B_R(\theta_R)\bigl[1-\epsilon_S(\theta_S)\bigr],
  \qquad
  S_{R+S}=S_0 .
\end{equation}
Since the signal contains no visible downstream charged particle, the
scintillator is used only as a veto. Neglecting accidental scintillator
activity, the signal yield is therefore unaffected by $\epsilon_S$.
For the charged-particle leakage component reaching the scintillator, this
corresponds to an additional suppression factor of approximately
$1-\epsilon_S=0.01$ relative to the 3-RPC-only configuration,
reducing the expected signal-region background from $\mathcal{O}(10^2)$ to $\mathcal{O}(1)$ events per simulated sample. The additional suppression reduces the expected residual background to the
few-event level in the present estimate and improves the projected coupling reach, as shown by the magenta curve in Fig.~\ref{fig:Uncertainty}. The magenta (hatched) band shows the corresponding degradation in the scintillator-assisted configuration, obtained by profiling the scintillator-efficiency nuisance together with the RPC calibration nuisance at the conservative $\delta\epsilon_{\rm RPC}=1\%$ level.

\subsection{Limit Extraction}

Detector-efficiency uncertainties are included through nuisance parameters in a profile-likelihood construction. 
For the RPC-only configuration, the likelihood is
\begin{equation}
  \mathcal{L}_R(n|\mu,\theta_R)
  = \mathrm{Pois}\!\bigl(n \mid \mu S_0 + B_R(\theta_R)\bigr)\,
  G(\theta_R),
\end{equation}
where $n$ is the observed event count in the signal region,
$\mu=(g_\phi/g_{\phi,\rm ref})^2$ is the signal-strength modifier,
$S_0$ is the expected signal yield at $g_{\phi,\rm ref}=10^{-6}$
after all nominal selections and scaled to the exposure considered,
$B_R(\theta_R)$ is the RPC-induced residual background defined in Appendix~\ref{sec:calibration}, and $G(\theta_R)$ denotes a unit-width Gaussian
constraint for $\theta_R$.
For the scintillator-assisted configuration, the likelihood becomes
\begin{equation}
  \mathcal{L}_{R+S}(n|\mu,\theta_R,\theta_S)
  =
  \mathrm{Pois}\!\bigl(n \mid \mu S_{R+S}
  + B_{R+S}(\theta_R,\theta_S)\bigr)
  G(\theta_R)G(\theta_S).
\end{equation}
The expected limits are evaluated with background-only Asimov data sets.

The test statistic is defined as
\begin{equation}
  q_\mu =
  -2\ln
  \frac{
  \mathcal{L}(\mu,\hat{\hat{\boldsymbol{\theta}}}_\mu)
  }{
  \mathcal{L}(\hat{\mu},\hat{\boldsymbol{\theta}})
  },
\end{equation}
where $\hat{\hat{\boldsymbol{\theta}}}_\mu$ and $(\hat{\mu},\hat{\boldsymbol{\theta}})$ denote the conditional and unconditional maximum-likelihood estimates, respectively. 
For the background-only Asimov data set, $\hat{\mu}=0$, and the one-sided 90\% CL upper limit is obtained from $q_{\mu_{90}}=2.7055$. 
The corresponding coupling limit is
\begin{equation}
  g_\phi^{90}
  =
  g_{\phi,\rm ref}\sqrt{\mu_{90}} .
\end{equation}

\begin{figure}[htbp]
    \centering
    \includegraphics[width=0.85\linewidth]{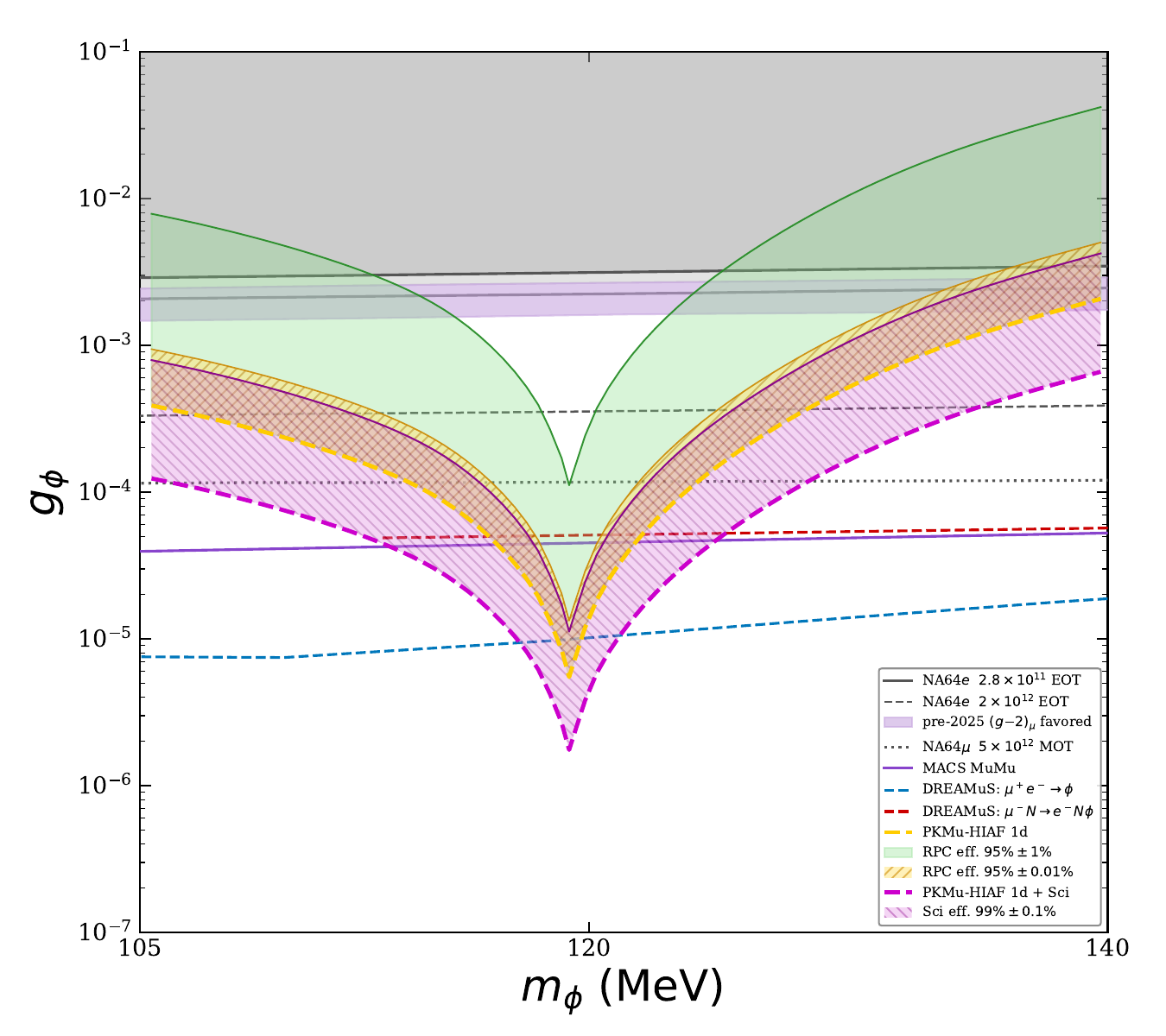}
        \caption{
        Projected 90\% CL expected upper limit on $g_\phi$ for one day of
        HIAF-PKMu running, including the impact of detector-efficiency
        uncertainties. The dashed yellow curve denotes the nominal RPC-only
        expected limit, and the dashed magenta curve denotes the nominal
        scintillator-assisted limit. The RPC-only degradation bands are shown for
        $\epsilon_{\rm RPC}=95\%\pm0.01\%$ in yellow and
        $\epsilon_{\rm RPC}=95\%\pm1\%$ in green. The magenta band shows the
        corresponding scintillator-assisted degradation, from the nominal +Sci
        limit to the profiled limit including
        $\epsilon_{\rm RPC}=95\%\pm1\%$ and $\epsilon_S=99\%\pm0.1\%$.
        In the displayed mass range, the lower and upper edges of each band
        correspond to the nominal and profiled limits.
        }
    \label{fig:Uncertainty}
\end{figure}

Figure~\ref{fig:Uncertainty} shows the effect of detector-efficiency uncertainties on the projected one-day limit. 
The RPC-only bands correspond to $\delta\epsilon_{\rm RPC}=0.01\%$ and $1\%$, while the scintillator-assisted band includes the conservative RPC uncertainty together with $\delta\epsilon_S=0.1\%$. 
The lower edge of each band gives the nominal statistical-only limit, already including the central detector efficiencies, and the upper edge gives the profiled limit after including the corresponding calibration uncertainties. 
When the background uncertainty dominates, the profiled limit approaches a systematic floor, explaining the weak luminosity dependence in the conservative $\delta\epsilon_{\rm RPC}=1\%$ case.

\section{Summary}

This work investigates the lepton-flavor-violating (LFV) interactions mediated by a new scalar boson $\phi$ that couples exclusively to electrons and muons, providing a clean and sensitive probe of physics beyond the Standard Model. We calculate the scattering amplitude and cross section using relativistic bound-state Dirac wave functions. Our numerical approach and detailed Monte Carlo simulations significantly improve the theoretical precision required for both current and future LFV searches by accurately accounting for target-dependent atomic motion effects. Furthermore, we have evaluated the experimental sensitivity of a concrete, fixed-target setup proposed for the HIAF facility to probe this $e-\mu$ LFV portal. Assuming no signal observation, the projected sensitivity can constrain the coupling strength to $g_{\phi} \sim 10^{-5}$ at the 90\% confidence level within the resonant mass window. This benchmark reach provides a substantially stronger constraint than the projected sensitivity of alternative searches such as NA64$\mu$, establishing the unique discovery potential of high-intensity muon beams at HIAF.

\acknowledgments

We are grateful to Prof. Luc Darm\'e (Universit\'e Claude Bernard Lyon)  and Prof. Huasheng Shao (LPTHE) for valuable theoretical discussions and support. We also thank Prof. Luc Darm\'e for supplying the numerical values of the f,g functions adopted in this work.

This work is supported in part by the National Natural Science Foundation of China under Grants No.~12325504.

\appendix

\section{Calibration}
\label{sec:calibration}

The downstream veto (Cut~2 in Table~\ref{tab:cutflow}) requires that none of the three downstream RPC planes records a hit. The background events surviving this cut consist predominantly of through-going charged particles that fail to register hits in all three planes due to finite RPC detection efficiency. To characterize these instrumental backgrounds, we calibrate the single-plane efficiencies $\epsilon_1$, $\epsilon_2$, $\epsilon_3$ using data-driven control regions.

For each event passing the upstream and fiducial selections, we record the hit pattern of the three downstream RPC planes,
\begin{equation}
  \mathbf{h} = (h_1, h_2, h_3), \qquad
  h_i = \begin{cases}
    1, & \mathrm{RPC}_i \mathrm{\ has\ a\ hit}, \\
    0, & \mathrm{otherwise}.
  \end{cases}
\end{equation}
The eight possible hit patterns ($000$ through $111$) define mutually exclusive event categories. The $000$ category corresponds to the downstream-veto signal-region topology. Assuming the per-plane RPC efficiencies are stable and the planes respond independently, the probability to observe a given pattern for a genuine through-going charged particle is
\begin{equation}
  P_{h_1h_2h_3}(\boldsymbol{\epsilon}) = \prod_{i=1}^{3} \epsilon_i^{h_i} (1 - \epsilon_i)^{1 - h_i}.
\end{equation}
Let $N_T$ be the total number of through-going charged background events entering the downstream stack. The expected yield in each event category is
\begin{equation}
  \lambda_{h_1h_2h_3} = N_T \, P_{h_1h_2h_3}(\boldsymbol{\epsilon}).
\end{equation}
With observed yields $n_{h_1h_2h_3}$, we construct a Poisson likelihood,
\begin{equation}
  \mathcal{L}_\mathrm{RPC}(N_T, \boldsymbol{\epsilon}) =
  \prod_{h_1,h_2,h_3=0}^{1}
  \mathrm{Pois}\!\bigl(n_{h_1h_2h_3} \mid \lambda_{h_1h_2h_3}\bigr).
  \label{eq:rpc_likelihood}
\end{equation}
Maximising Eq.~(\ref{eq:rpc_likelihood}) yields the best-fit efficiencies $\hat\epsilon_1$, $\hat\epsilon_2$, $\hat\epsilon_3$, the total normalisation $\hat N_T$, and their covariance matrix $\mathbf{V}_{\!\epsilon}$. The probability for a charged particle to escape detection by all three RPC planes, i.e.\ the instrumental background leakage into the signal region, is then
\begin{equation}
  q_R = \prod_{i=1}^{3} (1 - \epsilon_i).
\end{equation}
Its uncertainty is obtained by propagating $\mathbf{V}_{\!\epsilon}$ through the gradient of $q_R$,
\begin{equation}
  \sigma_{q_R}^2 = \sum_{i,j=1}^{3} 
  \frac{\partial q_R}{\partial \epsilon_i} 
  V_{\epsilon, ij} \,
  \frac{\partial q_R}{\partial \epsilon_j}.
\end{equation}

For the simulation study, we adopt a simplified model in which the dominant calibration bias is described by a single nuisance parameter $\theta_R$ constrained by a standard Gaussian prior, under the assumption that all RPC boards share a common efficiency.
\begin{equation}
  \epsilon_{\rm RPC}(\theta_R) = \epsilon_{\rm RPC,0} + \delta\epsilon_{\rm RPC}\,\theta_R,
\end{equation}
where $\epsilon_{\rm RPC,0}=95\%$ and $\delta\epsilon_{\rm RPC}$ is the absolute calibration uncertainty. The residual background yield after the downstream veto then scales as
\begin{equation}
  B_R(\theta_R) = B_0 \left[ \frac{1-\epsilon_{\rm RPC}(\theta_R)}{1-\epsilon_{\rm RPC,0}} \right]^3.
\end{equation}
This parameterisation conservatively assumes fully correlated systematic errors across the three RPC planes. The expected upper limit is obtained from the profile-likelihood ratio rather than by fixing the efficiency to a worst-case value. The degradation of the 1-day coupling limit after profiling the RPC-efficiency nuisance is shown in Fig.~\ref{fig:Uncertainty} as the yellow (hatched) band ($\delta\epsilon_{\rm RPC}=0.01\%$) and the green band ($\delta\epsilon_{\rm RPC}=1\%$).




\end{document}